      [1999/12/01 v1.4c Il Nuovo Cimento]
\documentclass{cimento}

\usepackage{graphicx}  
\newcommand{\open}{< \!\!\!) \;}
\title{Unpolarized and Polarized Fragmentation Functions}
\author{M.~Radici\from{ins:x}}
\instlist{\inst{ins:x} INFN - Sezione di Pavia - Pavia, Italy}
\PACSes{\PACSit{13.66.Bc}{Hadron production in $e^+e^-$ interactions}
\PACSit{13.87.Fh}{Fragmentation into hadrons}
\PACSit{13.88.+e}{Polarization in interactions and scattering}}

\begin{document}

\maketitle

\begin{abstract}
I give an overview of the present knowledge about nonperturbative functions parametrizing the fragmentation into one or two hadrons of (un)polarized light quarks in vacuum, including information on their transverse momentum dependence. 
\end{abstract}


\section{Introduction}
\label{sec:intro}

The fragmentation process describes the transition from a highly virtual parton $i$ at a scale $Q^2$ to one hadron $h$ carrying a fraction $z$ of its energy. The information is encoded in the fragmentation function $D_{1,i}^h(z,Q^2)$, which is a nonperturbative object since the hadronization itself is a soft, nonperturbative, process. As such, $D_{1,i}^h(z,Q^2)$ cannot be deduced from first principles but must be extracted from experiments. 

A large amount of data has been collected in the last 30 years by measuring hadron spectra in 
$e^+ e^-$ annihilations. Based on these data, several parametrizations of $D_{1,i}^h(z,Q^2)$ have been released.  More recently, new measurements in Semi-Inclusive Deep-Inelastic Scattering (SIDIS) and in hadronic ($p$-$p$ and $p$-$\bar{p}$) collisions were included in various fits. A very brief overview is given in Sec.~\ref{sec:1hFF} and the most updated parametrizations of  $D_{1,i}^h$ are compared in Sec.~\ref{sec:D1} (for brevity, only for light partons in vacuum; for a short review on medium modifications and heavier flavors, see Ref.~\cite{ref:ECT*proc}). 

The dependence of $D_{1,i}^h$ upon the transverse momentum $k_T$ of the fragmenting parton is basically unknown. Most of the phenomelogical studies are based on a simple flavor- and $z$-independent  Gaussian ansatz. But in several experimental results for hadron multiplicities the evidence emerges about transverse-momentum distributions depending on both energy and flavor of the detected hadron. This topic is directly addressed in Sec.~\ref{sec:kT}, and also in 
Sec.~\ref{sec:model} where the main three types of models of fragmentation functions are sketched. 

The above considerations apply also to polarized fragmentation functions, actually to the only one that has been parametrized so far: the Collins function. In fact, only its first $k_T$ moment could be extracted leaving the $k_T$ dependence fully unconstrained (see Sec.~\ref{sec:collins}). The Collins effect in spin asymmetries in SIDIS is one crucial tool to address the socalled transversity parton distribution~\cite{ref:collins}, a poorly known cornerstone in the knowledge of the (spin) partonic structure of the nucleon. That is why on one side models of the Collins function were developed and studied in detail (see Sec.~\ref{sec:spect} and \ref{sec:Artru}), and on the other side alternatives were considered in the "hunting for transversity". 

The most promising alternative is based on a spin asymmetry in SIDIS with two hadrons detected in the final jet. The corresponding Di-hadron Fragmentation Functions (DiFF) are encoded in functions like $D_{1,i}^{h_1 h_2}(z_1,z_2,M_h^2,Q^2)$, that must depend also on the invariant mass of the hadron pair, $M_h^2$; the latter represents a second natural scale in the fragmentation, with $M_h^2 \ll Q^2$~\cite{ref:noi02}. Further details and some first results are presented in 
Sec.~\ref{sec:DiFF}. 


\section{Single-hadron Fragmentation Functions}
\label{sec:1hFF}

In order to extract information on $D_{1,i}^h(z,Q^2)$ from data, the most suitable process is by far the electron-positron annihilation. Measuring the socalled scaled-energy distribution 
$(1 / \sigma_{{\rm tot}}) \frac{d\sigma^h}{dz}$  gives direct access at leading order (LO) in $\alpha_s$ to the fragmentation function summed over all active flavors~\cite{ref:Arleo105}. Well established factorization theorems~\cite{ref:Arleo1} allow to explore higher orders in terms of perturbatively calculable coefficient functions, that are known up to NNLÄO in the $\overline{MS}$ 
scheme~\cite{ref:Arleo108,ref:Arleo109}. A large amount of data has been collected in the last 30 years in a wide energy range, $12\leq Q \leq 200$ GeV and $0.005\leq z \leq 0.8$, and for various  hadron species: $\pi^\pm, K^\pm, K_s^0, p, \bar{p}, \Lambda, \bar{\Lambda}$ (see 
Ref.~\cite{ref:Arleo} and references therein, for a short review). Most experiments were able to disentangle the contribution of light quarks $(u,d,s)$ from $c$ and $b$ jets. In particular, the OPAL collaboration released also an analysis with full flavor separation~\cite{ref:Arleo67}. 

However, since at LO the $e^+ e^-$ annihilation leads to the back-to-back production of a quark and an antiquark jet, data only allow for the extraction of the flavor-inclusive fragmentation function 
$D_{1,q}^h + D_{1,\bar{q}}^h (\equiv D_{1,q}^h+D_{1,q}^{\bar{h}})$. Moreover, the gluon fragmentation function $D_{1,g}^h$ can only be extracted from 3-jet events that by construction appear at NLO; hence, it is weakly constrained. 

Fortunately, these drawbacks can be compensated by Semi-Inclusive Deep-Inelastic Scattering (SIDIS) data as well as by data on hadronic collisions. In the valence region 
$(x_{{\rm B}} \geq 0.1)$, quarks are produced more abundantly than antiquarks and individual 
$D_{1,q}^h$ or $D_{1,\bar{q}}^h$ can be independently extracted. In hadroproduction with $p$-$p$ and $p$-$\bar{p}$ collisions, the  $D_{1,g}^h$ can be directly addressed, particularly for 
$x_{{\rm B}} \ll 1$ or, equivalently, for hadron transverse momenta $P_\perp$ small with respect to the center-of-mass (cm) energy $\sqrt{s}$ available in the collision. Moreover, since 
$x_{{\rm B}} = {\cal O} (P_\perp / z)$ the much larger parton densities of the projectile/target at small  $x_{{\rm B}}$ allow to probe the fragmentation functions at large $z$ $(> 0.7)$, complementing the information extracted in $e^+ e^-$ annihilations. 

SIDIS data have been collected in the last 15 years for both unidentified $(h^\pm)$ and identified charged hadrons ($\pi^\pm, K^\pm$, and also $\Lambda, \bar{\Lambda}$) mostly in $e^- p$ collisions at HERA (H1~\cite{ref:Arleo78,ref:Arleo79,ref:Arleo80}, HERMES~\cite{ref:Arleo81}, and ZEUS~\cite{ref:Arleo82,ref:Arleo83,ref:Arleo84} collaborations), also at CERN with muonic (anti)neutrino beams (NOMAD~\cite{ref:Arleo85}). The explored kinematical range, 
$1 \leq Q \leq 100$ GeV and $0.1 \leq z < 1$, significantly enlarges the phase space available to $D_{1,i}^h(z,Q^2)$, since the hard scale is not constrained at the cm energy as in 
$e^+ e^-$ annihilations, $Q = \sqrt{s} / 2$. By analyzing scaled-energy distributions in the Breit frame for events in the kinematical current region at $z = P_h / (Q/2)$, it was possible to compare the results with the corresponding $e^+ e^-$ ones and to successfully test the universality of fragmentation functions~\cite{ref:Arleo80}. 

Hadron spectra in hadronic collisions appeared more recently, thanks to high-precision  $p$-$p$ measurements at RHIC (BRAHMS~\cite{ref:Arleo88}, PHENIX~\cite{ref:Arleo89,ref:Arleo90}, and STAR~\cite{ref:Arleo91,ref:Arleo92,ref:Arleo93} collaborations) and $p$-$\bar{p}$ ones by 
CDF~\cite{ref:Arleo95,ref:Arleo102} at the Tevatron. At RHIC, $P_\perp$ distributions of $\pi$'s, 
$K$'s, and protons, were measured up to 10 GeV at mid-to-large rapidities (for $\pi^0$, also up to 20 GeV by PHENIX~\cite{ref:Arleo90}), as well as puzzling data for $\Lambda, \bar{\Lambda}$, by the STAR collaboration~\cite{ref:Arleo93}. 


\subsection{Unpolarized fragmentation}
\label{sec:D1}

The year 2007 represents a sort of turning point for the phenomenological work about extraction of 
$D_{1,i}^h(z,Q^2)$ from experiments. All parametrizations released before this date are based on 
$e^+ e^-$ data only, they suffer from large uncertainties at large $z$ and $Q^2$, and fail to reproduce the scaling violations diplayed by SIDIS data reported by the H1 collaboration~\cite{ref:Arleo80}. 

On year 2007, two parametrizations (HKNS~\cite{ref:Arleo10} and 
DSS~\cite{ref:Arleo11,ref:Arleo12}, including also~\cite{ref:Arleo13}), followed by AKK08~\cite{ref:Arleo14} one year later, have been released which include also data from SIDIS and hadronic collisions, and show an error analysis in the fit. Their main features are listed in Tab.~\ref{tab:D1param}. 

\begin{table}
  \caption{Main features of global fit analyses DSS~\cite{ref:Arleo11,ref:Arleo12,ref:Arleo13}, 
  HKNS~\cite{ref:Arleo10}, and AKK08~\cite{ref:Arleo14}: data sample, kinematic range covered, technique for error analysis.}
  \label{tab:D1param}
  \begin{tabular}{rcl}
    \hline
     DSS      & HKNS & AKK08    \\
    \hline
    \hline
      $e^+ e^-$,  SIDIS, $pp$  & $e^+ e^-$ &  $e^+ e^-$,  $pp$,  $p\bar{p}$ \\
      $0.05\leq z$ ,  $1\leq Q^2\leq 10^5$ GeV$^2$    &  $0.01\leq z$ ,  $1\leq Q^2\leq 10^8$ GeV$^2$ & $0.05\leq z$ ,  $2\leq Q^2\leq 4\times 10^4$ GeV$^2$ \\
    \hline
      Lagrange multipliers   & Hessian errors  & in progress \\
    \hline
  \end{tabular}
\end{table}

The scaling violations are described by solving evolution equations at NLO with different techniques and constraining them to reproduce the longitudinal momentum sum rule. SU(2) isospin symmetry is assumed for the unfavoured (sea) channel at the starting scale; AKK08 and HKNS assume it also for the favoured one. AKK08 further includes the resummation of leading (LL) and next-to-leading (NLL) logarithms for $z\to 1$ both in the evolution equations and in the coefficient functions of the factorization formula, somewhat confusing the comparison with the other fixed-order extractions. 

In fact, AKK08 and DSS produce very similar results but at large $z$~\cite{ref:Arleo11}, where the effect of large logarithms is dominant; HKNS gives a doubtful softer gluon $D_{1,g}^{\pi^\pm}$ because it lacks the constraint from RHIC $pp$ data~\cite{ref:Arleo10}. Remarkably, all sets fail to reproduce the STAR data for $\Lambda, \bar{\Lambda}$ production in $pp$ collisions by almost one order of magnitude~\cite{ref:Arleo93}. 


\subsection{Tranverse-momentum dependence}
\label{sec:kT}

The completely unknown dependence upon the transverse momentum of partons is usually parametrized in terms of a Gaussian ansatz. In SIDIS, the Gaussian width is fixed, for example, by reproducing the data for the average transverse momentum squared  
$\langle {\bf P}_{h\perp}^2\rangle$ of final hadron $h$ with respect to the virtual photon direction in the lab~\cite{ref:Collins06}, because 
\begin{equation}
\langle {\bf P}_{h\perp}^2\rangle = z^2 \langle {\bf p}_\perp^2\rangle + \langle {\bf K}_T^2\rangle \; , 
\quad  {\bf K}_T = - z  {\bf k}_T \; , 
\label{eq:Tmom}
\end{equation}
where $\bf{p}_\perp$ refers to the initial parton considered again in the lab frame, while ${\bf K}_T$ is the hadron transverse momentum with respect to the direction of the fragmenting parton (approximately, the jet axis) and, viceversa, ${\bf k}_T$ refers to the fragmenting parton with respect to the final hadron. 

A new combined analysis of recent SIDIS data (including azimuthally asymmetric $\cos\phi$ and 
$\cos2\phi$ modulations in the cross section) and Drell-Yan data has lead to a new parametrization (see Ref.~\cite{ref:STM10} and references therein). The best fit gives 
$\langle {\bf p}_\perp^2\rangle = 0.38 \pm 0.06 , \; \langle {\bf K}_T^2\rangle = 0.16 \pm 0.01$ 
GeV$^2$, with a linear dependence in the cm energy $s$ which broadens the distributions for increasing energy. 

\begin{figure}
\begin{center}
\includegraphics[width=0.7\textwidth]{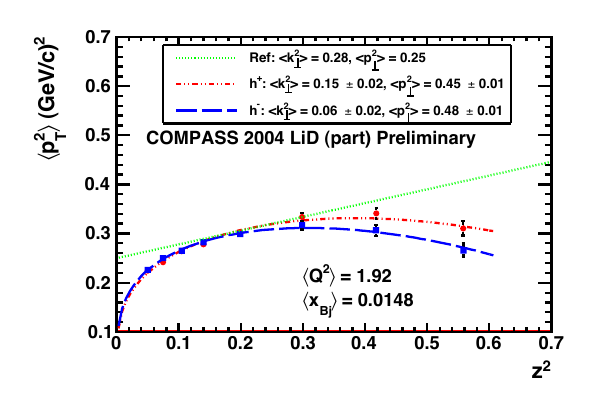}
\end{center}
\vspace{-0.5cm}
\caption{\label{fig:compass} The COMPASS 2004 data for $\langle {\bf P}_{h\perp}^2\rangle$ as 
a function of $z^2$~\cite{ref:Raj10}. Lower (higher) curves refer to fits for negative (positive) charged final hadrons. Straight (green) line obtained with constant Gaussian widths.}
\end{figure}

However, there are several indications in SIDIS measurements that the Gaussian widths should depend at least on $z$ and on the flavor content of the final hadrons. For example, in 
Fig.~\ref{fig:compass} the COMPASS 2004 data for 
$\langle {\bf P}_{h\perp}^2\rangle$~\cite{ref:Raj10} show a clear dependence on $z^2$ and on the hadron charge, the straight line being obtained with constant 
$\langle {\bf p}_\perp^2\rangle, \, \langle {\bf K}_T^2\rangle$, and fitting only few data points. 

\begin{figure}
\begin{center}
\includegraphics[width=0.5\textwidth]{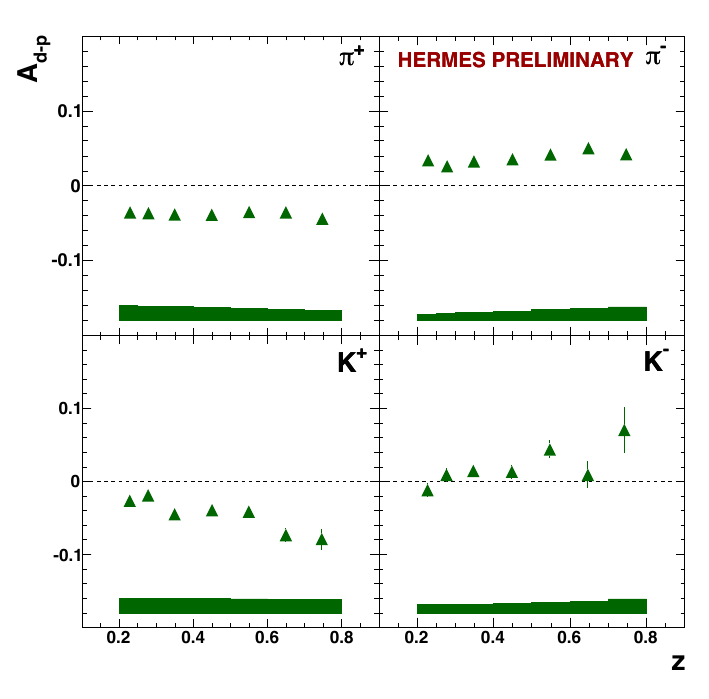}
\end{center}
\vspace{-0.5cm}
\caption{\label{fig:hermes} The HERMES data for the multiplicity asymmetry between deuteron and proton targets as a function of $z$~\cite{ref:JooDIS11}. Higher (lower) panels for pions (kaons), left (right) columns for positive (negative) charges.}
\end{figure}

Similarly, in Fig.~\ref{fig:hermes} the asymmetry between deuteron and proton targets of multiplicities for pions and kaons are displayed as functions of $z$ for recent HERMES 
data~\cite{ref:JooDIS11}. Nonvanishing (sometimes, large) asymmetries indicate the marked sensitivity of final-state distributions to the different flavor content of different targets. 

Finally, in Ref.~\cite{ref:RogAy11} a first attempt in describing the evolution of transverse-momentum dependent (TMD) nonperturbative functions, either initial distributions or final fragmentations, was put forward in the context of a proper factorization theorem. It turns out that even at LO, for the test case $D_{1,u}^{\pi^+}$ of interest here, the ${\bf K}_T$ distribution strongly depends on the hard scale $Q^2$ even at very low values of $|{\bf K}_T|$, getting broader and broader with increasing $Q^2$. 


\subsection{Polarized fragmentation}
\label{sec:collins}

There is only one polarized fragmentation function that has been extracted from experimental data so far: it is the socalled Collins function $H_{1}^{\perp}$~\cite{ref:collins}. It is related to the probability density of having a distorted ${\bf P}_{h\perp}$ distribution of the final hadron $h$ depending on the direction of the transverse polarization ${\bf S}_q$ of the fragmenting quark via the spin-orbit effect ${\bf S}_q \cdot \hat{\bf k} \times {\bf P}_h$, with $\hat{\bf k}$ pointing in the direction of the jet axis. 

It can appear in $e^+ e^- \to q^\uparrow q^\downarrow \to h^+ h^- X$ events or, more interestingly, in SIDIS on transversely polarized targets. In fact, a specific azimuthally asymmetric modulation of the leading-twist SIDIS cross section contains the convolution $h_1^q \otimes H_{1,q}^{\perp\, h}$ on the transverse momenta of the initial and final quarks, where $h_1$ is the socalled transversity parton distribution, a poorly known cornerstone in the construction of the (spin) partonic structure of the nucleon (for a review, see for example Ref.~\cite{ref:BRD02}). 

Both azimuthal asymmetries in $e^+ e^-$ and in SIDIS have been measured by the 
BELLE~\cite{ref:Abe06} and HERMES and COMPASS~\cite{ref:Airap05,ref:Ageev07} collaborations, respectively, and also later with increased 
statistics~\cite{ref:Seidl08,ref:Dief07,ref:Alek09}. A simultaneous fit of the three data sets made it possible for the first time to extract a parametrization for the transversity 
$h_1$~\cite{ref:TO07,ref:TO09}. The results are presently limited to the valence $u, d,$ quarks because of the limited kinematical range covered by the experiments. As for the fragmentation, we speak of favoured $(u\to \pi^+, \, d \to \pi^-)$ and unfavoured $(u\to \pi^-, \, d \to \pi^+)$ channels with the surprising and interesting findings that 
$H_{1,u}^{\perp\, \pi^-} \approx - H_{1,u}^{\perp\, \pi^+}$~\cite{ref:Airap05}. 

As prevously said, the Collins function is paired to the transversity in a convolution on quark transverse momenta. Hence, the knowledge of the whole $H_1^{\perp} (z, {\bf K}_T)$ function is crucial to unravel the convolution. However, from the expression of the azimuthal asymmetry in 
$e^+ e^-$ only a ${\bf K}_T$-integrated moment of $H_1^{\perp}$ can be isolated, leaving its 
${\bf K}_T$ dependence unconstrained. The latter has been parametrized similarly to $D_1$, i.e. with a Gaussian ansatz independent of kinematics and of the involved flavor. Therefore, the extraction of $h_1$ in Ref.~\cite{ref:TO07,ref:TO09} is affected by a model dependence. Moreover, the azimuthal asymmetries in $e^+ e^-$ and in SIDIS were measured at two very different scales, $Q^2=100$ and $2.5$ GeV$^2$ respectively, but evolution effects in the ${\bf K}_T$ dependence were neglected. Partial results on the evolution properties of the Collins function have been recently published~\cite{ref:Kang11}, but it is fair to say that a full treatment of TMD evolution in the Collins effect is still missing. 


\section{Models}
\label{sec:model}

Since the extraction of fragmentation functions from experimental data is affected by large uncertainties, as we have seen about the Collins function and, more generally, about the ${\bf K}_T$ dependence acquired by hadrons during the fragmentation, it is desirable that this phenomenology is supported by model speculations. In the following, we sketch three main classes of models that appeared in the recent literature. 


\begin{figure}
\begin{center}
\includegraphics[width=0.5\textwidth]{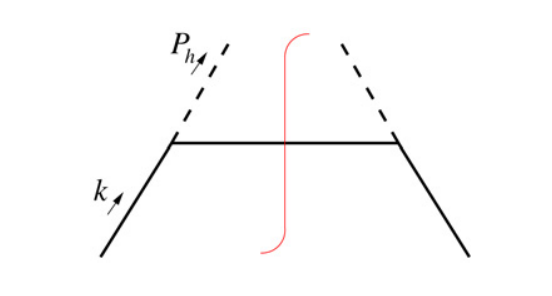}
\end{center}
\vspace{-0.5cm}
\caption{\label{fig:spect} The spectator approximation for a parton with momentum $k$ fragmenting into a detected hadron with momentum $P_h$.}
\end{figure}

\subsection{Spectator approximation}
\label{sec:spect}

The spectator approximation amounts to describe the fragmentation as the decay of a parton with momentum $k$ into the observed hadron $h$ with momentum $P_h$ leaving a residual system in an on-shell state with momentum $k-P_h$ (see the diagram in Fig.~\ref{fig:spect}). The latter condition grants that most of the calculations can be performed analytically, including the expression for the off-shellness $k^2(z)$ of the fragmenting parton. The drawback is that only the favoured channel can be taken into account. 

For the typical $u\to \pi^+$ channel, two main choices have been adopted in the literature for the quark-pion-spectator vertex: the pseudoscalar coupling 
$g_{\pi q} \gamma_5$~\cite{ref:Jak97,ref:Ale01,ref:Ale05,ref:Ale08,ref:Gamb03} and the pseudovector coupling 
$g_{\pi q} \gamma_5 \gamma_\mu P_h^\mu$~\cite{ref:Ale02,ref:Ale03,ref:Ale05}. In all cases the coupling was assumed to be point-like except in Refs.~\cite{ref:Gamb03,ref:Ale08}, where a gaussian form factor was used with a $z$-dependent cut-off. 

\begin{figure}
\begin{center}
\includegraphics[width=0.6\textwidth]{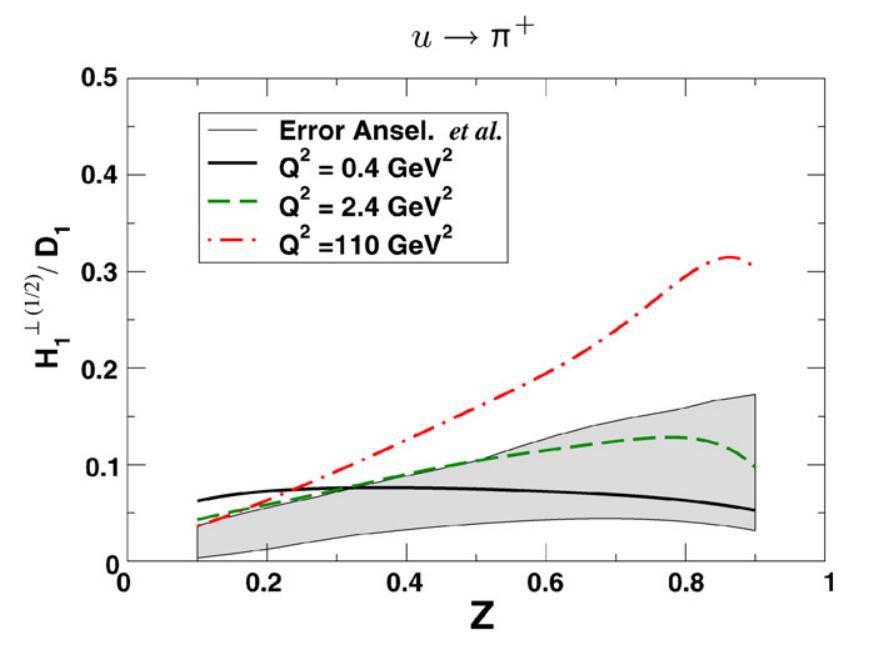}
\end{center}
\vspace{-0.5cm}
\caption{\label{fig:spectCollins} The spectator model result for 
$H_{1,u}^{\perp\,\pi^+ (1/2)}(z) / D_{1,u}^{\pi^+}(z)$ from Ref.~\cite{ref:Ale08} (see text). Solid (black), dashed (green), and dot-dashed (red) lines for $Q^2=0.4, 2.4, 110$ GeV$^2$, respectively. Uncertainty band for the phenomenological extraction of Ref.~\cite{ref:TO07}.}
\end{figure}

Complicated objects like the Collins function appear if there are nonvanishing interference diagrams involving different channels. In the spectator approximation, these final-state interactions can be achieved by adding to the left or right side of the diagram in Fig.~\ref{fig:spect} insertions involving pions and/or gluons. As an example, in Fig.~\ref{fig:spectCollins} the ${\bf K}_T$- integrated 
$\textstyle{\frac{1}{2}}$-moment $H_{1,u}^{\perp\, \pi^+ (1/2)}$ (normalized to $D_{1,u}^{\pi^+}$) from Ref.~\cite{ref:Ale08} is plotted as a function of $z$ for three different hard scales and compared with the parametrization of Ref.~\cite{ref:TO07}, whose statistical error is represented by the uncertainty band. The spectator results were obtained using a pseudoscalar $q\pi $ coupling and gluon insertions. The model parameters were fixed by reproducing the unpolarized $D_1$ at the lowest available $Q^2=0.4$ GeV$^2$, as it was extracted from $e^+ e^-$ data in 
Ref.~\cite{ref:Kretz00}. Since the parametrization of $H_1^\perp$ was performed using SIDIS data for the Collins effect at $Q^2=2.5$ GeV$^2$, the band in Fig.~\ref{fig:spectCollins} should be compared with the dashed (green) line, showing a substantial agreement with the spectator model. 


\begin{figure}
\begin{center}
\includegraphics[width=0.5\textwidth]{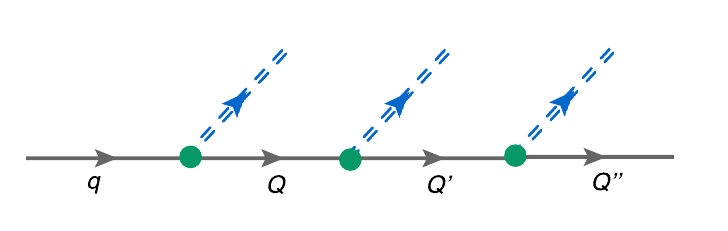}
\end{center}
\vspace{-0.5cm}
\caption{\label{fig:NJLjet} Recursive quark-meson splitting as the elementary mechanism of fragmentation in the Nambu-Jona Lasinio jet model.}
\end{figure}

\subsection{Nambu-Jona Lasinio jet model}
\label{sec:NJLjet}

In the Nambu-Jona Lasinio (NJL) jet model~\cite{ref:Ito09}, the fragmentation is represented as a recursive process depicted in Fig.~\ref{fig:NJLjet}. The $D_{1,q}^m (z)$ can be obtained by solving a set of coupled integral equations based on the following product ansatz:
\begin{equation}
D_{1,q}^m (z) dz = d_q^m (z) dz + \sum_Q \left[ d_q^Q \otimes D_Q^m \right] (z) \; , 
\label{eq:NJLjet}
\end{equation}
where the elementary fragmentation function $d_q^m$ of a quark $q$ in the meson $m$ describes each splitting step in Fig.~\ref{fig:NJLjet}, and the sum runs upon all allowed intermediate states 
$Q$ in the cascade. In each splitting, the $d_q^m$ depends on the quark-meson coupling 
$g_{qmQ}$, which is determined from the residue (at the pole of the meson mass) in the 
quark-antiquark T matrix~\cite{ref:Matev11K}. 

However, the above framework is justified only in the Bjorken limit where the quark initiating the cascade has an infinite momentum and produces an infinite number of hadrons; only in this limit the momentum sum rule is satisfied. Moreover, the generalization to fragmentation into baryons is not trivial. Finally, solving the coupled integral equations is sometimes a heavy computational task. For all these reasons, the quark-cascade description of the fragmentation has been approached using the Monte Carlo technique~\cite{ref:Matev11}. In this context, the fragmentation function 
$D_q^h(z) \Delta z$ for a hadron $h$ with momentum fraction in the range $[z, z+\Delta z]$, is deduced by calculating the average number of hadrons of type $h$ produced in the cascade depicted in Fig.~\ref{fig:NJLjet} for a predefined number of steps $N_{{\rm step}}$. Each step of the cascade is randomly sampled using the $d_q^h$ calculated in the NJL jet model, and the entire cascade is simulated $N_{{\rm sim}}$ times with $N_{{\rm sim}}$ large enough to stabilize the average. When the hadron $h$ is a baryon, the internediate states $Q$ are described in the framework of the spectator scalar diquark model~\cite{ref:Matev11}. 

\begin{figure}
\begin{center}
\includegraphics[width=0.9\textwidth]{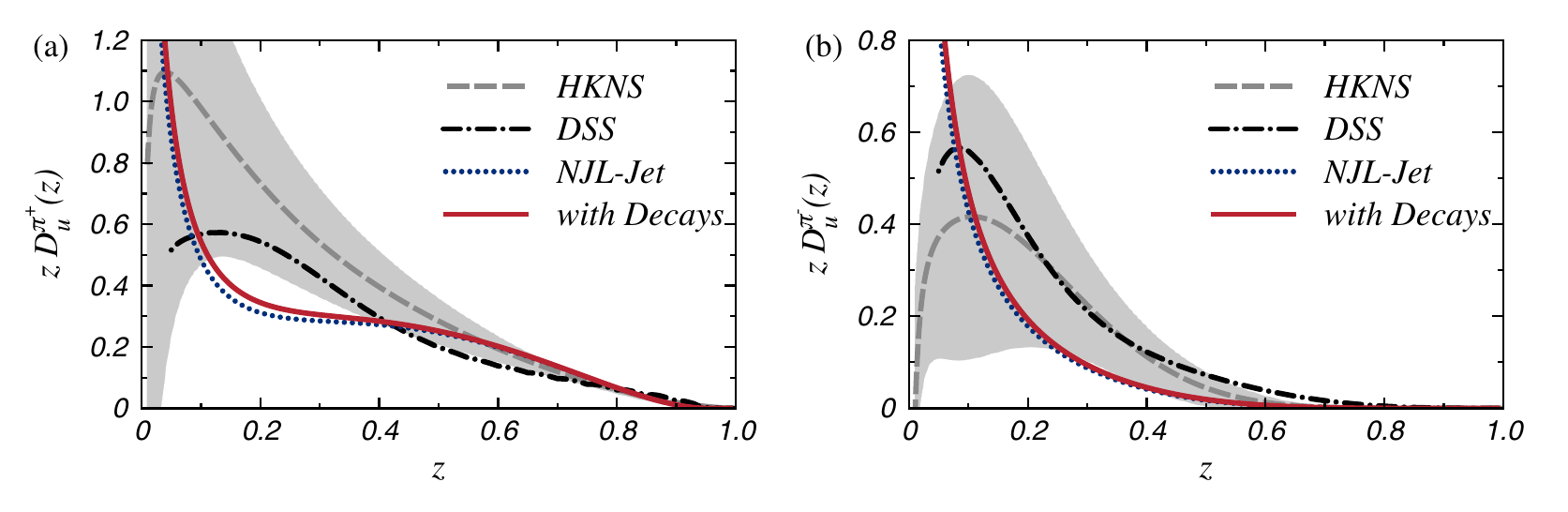}
\end{center}
\vspace{-0.5cm}
\caption{\label{fig:NJLu} The $z D_{1,u}^h (z)$ for $h=\pi^+$ (left panel) and $h=\pi^-$ (right panel). Dashed and dot-dashed lines for the parametrizations of HKNS~\cite{ref:Arleo10} and 
DSS~\cite{ref:Arleo11}, the uncertainty bands referring to HKNS. Dotted line for the result of the NJL jet model, solid (red) line  includes in the cascade also intermediate vector 
mesons~\cite{ref:Matev11}.}
\end{figure}

In Fig.~\ref{fig:NJLu}, the $z D_{1,u}^h (z)$ is shown for the favoured ($h=\pi^+$, left panel) and unfavoured channels ($h=\pi^-$, right panel). The dashed and dot-dashed lines represent the phenomenological parametrizations of HKNS~\cite{ref:Arleo10} and DSS~\cite{ref:Arleo11} (see 
Tab.~\ref{tab:D1param}), the uncertainty bands referring to HKNS. The result of the NJL jet model is represented by the dotted line, while the solid (red) one includes in the cascade also intermediate vector mesons, further decaying into the observed $\pi^\pm$. For sake of consistency with the empirical parametrizations, the NJL jet model results are evolved at NLO from $Q_0^2=0.2$ to $Q^2=4$ GeV$^2$. The Lepage-Brodsky scheme adopted in regularizing the calculation of loop diagrams reduces the available range in $z$ depending on the hadron type $h$. For 
$z\to z_{{\rm min}}(h)$ the Monte Carlo result for $z D_1(z)$ tends to a constant which becomes larger for increasing $Q^2$, from which the apparently divergent behaviour for small $z$. We refer the interested reader to Ref.~\cite{ref:Matev11} for the results about other channels. 


\begin{figure}
\begin{center}
\includegraphics[width=0.6\textwidth]{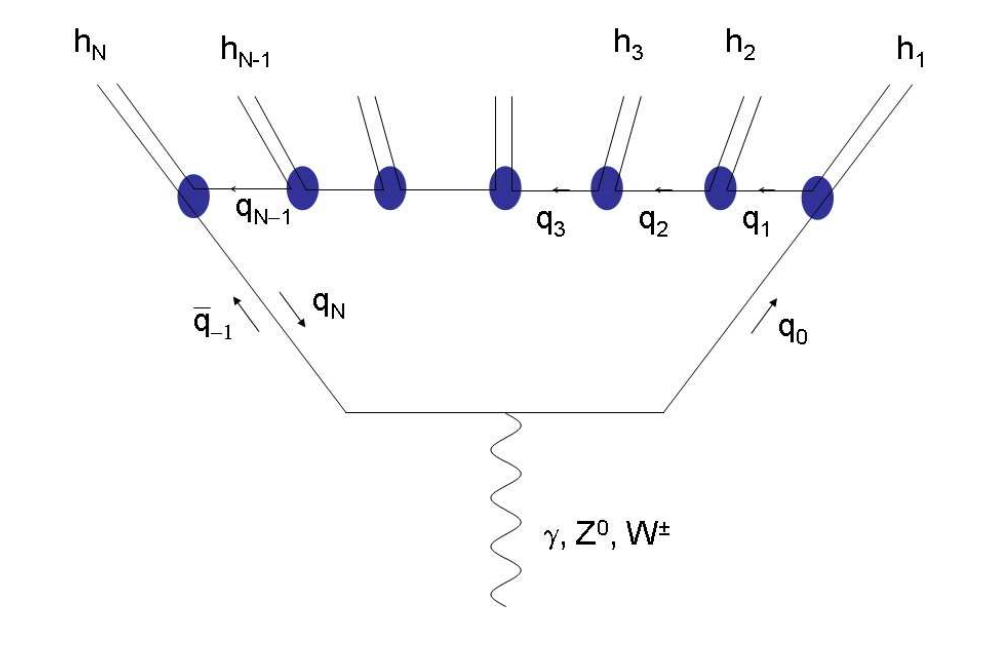}
\end{center}
\vspace{-0.5cm}
\caption{\label{fig:Artru} The process $e^+ e^- \to q_0 \bar{q}_{-1} \to h_1+h_2+\dots +h_N$ as a recursive $q\to hq'$ splitting.}
\end{figure}

\subsection{Recursive model}
\label{sec:Artru}

Present Monte Carlo event generators of quark and gluon jets do not include spin in the elementary degrees of freedom. Therefore, in order to have a guide in studying azimuthal asymmetries like the Collins effect, a new quantum approach to polarized quark fragmentation was suggested in 
Ref.~\cite{ref:Artru10}. As an example, the production of $N$ pseudo-scalar mesons $h_1,h_2,\ldots,h_N$, in $e^+ e^-$ annihilation is depicted in Fig.~\ref{fig:Artru} via the elementary annihilation $q_0 \bar{q}_{-1}$, where in each step $i$ of the chain, from right to left, we have 
$q_{i-1}=h_i+q_i$ for $i=1,..,N$, and $q_N=-\bar{q}_{-1}$. 

Considering the upper part of Fig.~\ref{fig:Artru} as a Feynman amplitude with spinors, vertices and propagators, it is possible to study its dependence on spin by introducing some {\it ad hoc} simplifications. In Ref.~\cite{ref:Artru10}, this amplitude was estimated in a socalled multiperipheral model by using Pauli spinors and matrices in the vertices, and by approximating each intermediate fermion propagator at step $i$ with an expression similar to the meson-nucleon scattering amplitude,
\begin{equation}
\Delta_i \approx \exp [-b  {\bf h}_{iT}^2/2]\, \left[ \mu ({\bf h}_{iT}^2) + {\rm i} {\bf \sigma} \cdot 
\hat{\bf z} \times {\bf h}_{iT} \right]  \; , 
\label{eq:Artru}
\end{equation}
i.e. with a non-spin-flip complex function $\mu$ and a spin-flip part, $b$ being some free parameter. These prescriptions can be shown to respect invariance under all "good" transformations like rotations, boosts, and parity, all considered with respect to the jet axis $\hat{\bf z}$. 

If ${\rm Im}(\mu) \ne 0$, this imaginary part can be shown to act as a source of transverse polarization at step $i$ even if the quark was unpolarized or longitudinally polarized at step 
$i-1$~\cite{ref:Artru10}. This means also that during the cascade the helicity of a quark can be partly converted to its transversity or viceversa.  As a consequence, if ${\rm Im}(\mu) \ne 0$ one can have for $N=1$ a Collins effect ${\bf S}_1 \cdot \hat{\bf z} \times {\bf h}_{1T}$, and for $N=2$ an iterated Collins effect with alternate sign, which could explain the experimental findings 
$H_{1}^{\perp\, {\rm unf}} \approx - H_{1}^{\perp\, {\rm fav}}$ described in 
Sec.~\ref{sec:collins}~\cite{ref:Airap05}. This result confirms the outcome of the Lund $^3P_0$ string mechanism~\cite{ref:Lund}. But in addition it contains the three-particle correlation 
$\hat{\bf z} \cdot {\bf h}_{2T} \times {\bf h}_{1T}$ named jet handedness~\cite{ref:jethand}, which is interpreted as a two-step mechanism: at $i=1$, a transverse polarization 
${\bf S}_{1T} \parallel {\bf h}_{1T}$ is generated from the helicity ${\bf S}_{0z}$ of previous step; at $i=2$, a Collins effect takes place as $\hat{\bf z} \cdot {\bf h}_{2T} \times {\bf S}_{1T}$, which coincides with the jet handedness. 

Further work is needed to promote the multiperipheral model of Ref.~\cite{ref:Artru10} to a realistic Monte Carlo event generator. For example, one should include antiquarks in the fragmentation cascade, or explore the interference of the amplitude in Fig.~\ref{fig:Artru} with diagrams showing differently ordered $N$ hadrons. Preliminary experimental results already appeared for $K^-$ SIDIS production by the HERMES collaboration (an almost vanishing Collins 
effect~\cite{ref:Airap09} and a large $\cos2\phi$ asymmetry in the unpolarized cross 
section~\cite{ref:Gunarhere}) that cannot be easily accommodated in the multiperipheral model in its present version. 


\begin{figure}
\begin{center}
\includegraphics[width=0.55\textwidth]{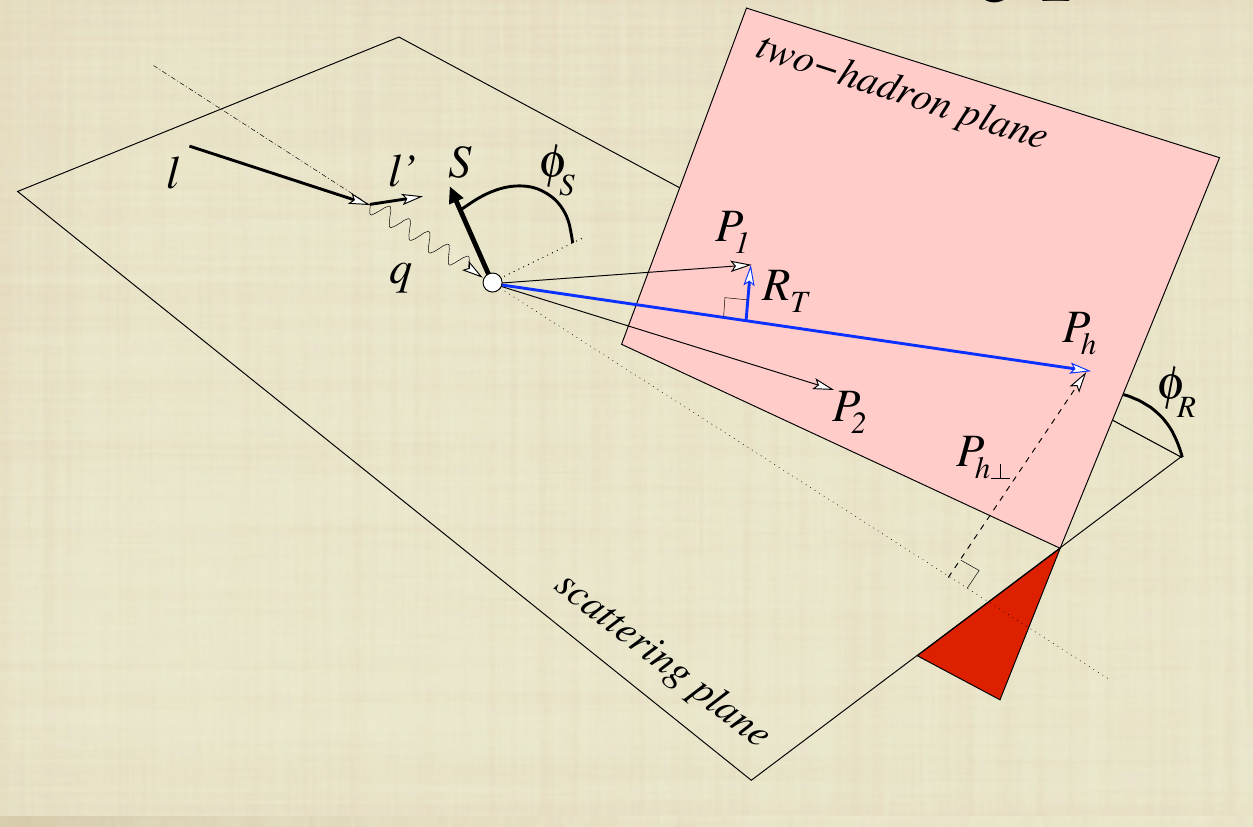}\hspace{.2cm}
\includegraphics[width=0.4\textwidth]{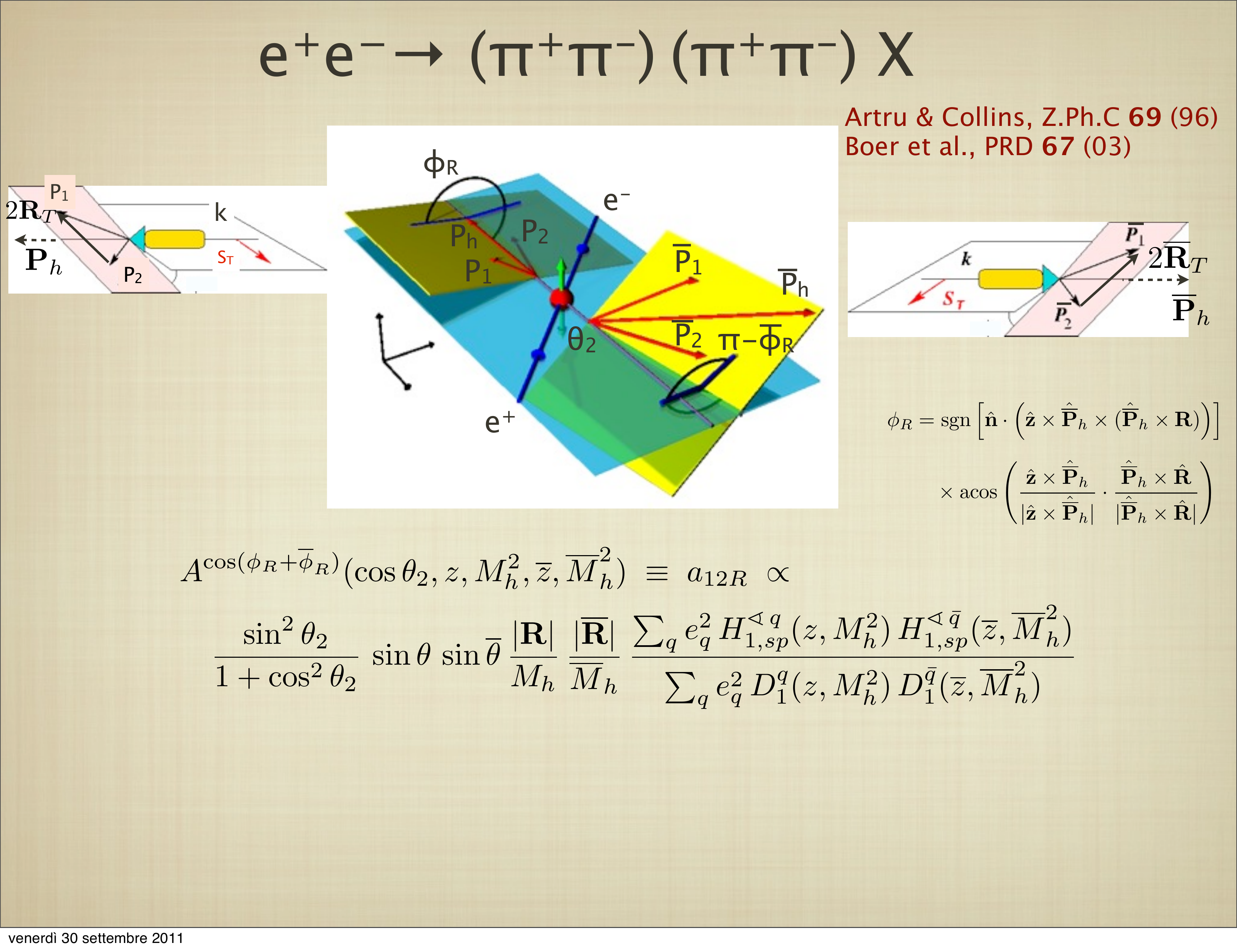}
\end{center}
\vspace{-0.5cm}
\caption{\label{fig:DiFFkin} The kinematics for the SIDIS process $e p^\uparrow \to e' (h_1 h_2) X$ (left panel) and for the process $e^+ e^- \to (h_1 h_2) (h_1' h_2') X$ (right panel, adapted from 
Ref.~\cite{ref:BELLE}.}
\end{figure}

\section{Di-hadron Fragmentation Functions}
\label{sec:DiFF}

As already sketched in Sec.~\ref{sec:collins}, the extraction of the transversity parton distribution via the Collins effect suffers from several uncertainties and model dependencies, mostly related to the need of dealing with TMD objects. A complementary approach is provided by the semi-inclusive process $e p^\uparrow \to e' (h_1 h_2) X$ where two unpolarized hadrons with momenta $P_1$ and $P_2$ emerge from the fragmentation of the same quark. The kinematics is similar to the 
single-hadron SIDIS except for the final state, where the hadron pair carries a fractional energy $z=z_1+z_2$ with a total momentum $P_h=P_1+P_2$ and a relative momentum $R=(P_1-P_2)/2$ (see Fig.~\ref{fig:DiFFkin}, left panel). The underlying mechanism is related to 
${\bf S}_q \cdot {\bf P}_h \times {\bf R}$: the transverse polarization of the fragmenting quark $q$ is transferred to the relative orbital angular momentum of the hadron pair. Contrary to the Collins effect, this mechanism survives after integrating away the transverse momentum of each particle and can be analyzed in the collinear factorization scheme~\cite{ref:noi02}. The probabilistic weight for this to happen is represented by the polarized dihadron fragmentation function 
$H_{1,q}^{\open\,h_1 h_2} (z,M_h^2, Q^2)$, where $P_h^2=M_h^2 \ll Q^2$ is the pair invariant mass and represents a new soft scale in the process. It is easy to show that 
$|{\bf R}|^2$ is a linear function of $M_h^2$~\cite{ref:noi03}. 

Dihadron Fragmentation Functions (DiFF) were introduced in Ref.~\cite{ref:Venez78} and studied for the polarized case in Refs.~\cite{ref:CHL94,ref:JJT98,ref:ArtCol96}. The decomposition of the SIDIS cross section in terms of parton distributions and DiFF was carried out at leading twist in 
Ref.~\cite{ref:noi00} and to sub-leading twist in Ref.~\cite{ref:noi04}. Since 
${\bf R}_T = {\bf R} \sin\theta$, where in cm frame of the hadron pair $\theta$ is the angle between $P_1$ and the direction of $P_h$ in the lab frame~\cite{ref:noi03}, the leading-twist cross section shows an azimuthally asymmetric modulation proportional to $\sin (\phi_R+\phi_S) \sin\theta$, where $\phi_R, \, \phi_S$, are defined in Fig.~\ref{fig:DiFFkin}. The proportionality coefficient contains the product 
$h_1^q \, H_{1,q}^{\open\,h_1 h_2}$~\cite{ref:JJT98,ref:noi02,ref:noi03,ref:noi00,ref:noi00bis}: the advantage of working in collinear factorization scheme reflects in a very simple relation with no convolution on transverse momenta, as in the case of the Collins effect. 

The $H_{1,q}^{\open\,h_1 h_2}$ is sensitive to the interference between the fragmentation amplitudes into hadron pairs in relative $s$ wave and in relative $p$ wave~\cite{ref:noi03}. Intuitively, if the fragmenting quark is moving along $\hat{\bf z}$ and is polarized along 
$\hat{\bf y}$, a positive $H_{1,q}^{\open\,h_1 h_2}$ means that $h_1$ is preferentially emitted along $-\hat{\bf x}$ and $h_2$ along $\hat{\bf x}$. The corresponding unpolarized partner 
$D_{1,q}^{h_1 h_2}$ is averaged over quark polarization and hadron pair orientation. Similarly to the single-hadron SIDIS, the unknown DiFF must be independently determined from $e^+ e^-$ annihilation producing, in this case, two hadron pairs (see right panel of Fig.~\ref{fig:DiFFkin}, adapted from Ref.~\cite{ref:BELLE}). The relevant signal is similar to that of the Collins function, except that each transverse polarization of the $q\bar{q}$ pair is now correlated to the azimuthal orientations $\phi_R, \phi_{\bar{R}}$, of the planes formed by the momenta of the corresponding hadron pairs, suggesting that $H_{1}^{\open}$ is related to the above mentioned concept of jet handedness~\cite{ref:jethand,ref:e+e-03}. In the leading-twist cross section, this correlation shows up as a modulation proportional to $\cos (\phi_R + \phi_{\bar{R}})$~\cite{ref:e+e-03}. 

The $A_{UT}^{\sin (\phi_R+\phi_S) \sin\theta}$ spin asymmetry in SIDIS production of a $\pi^+ \pi^-$ pair off a transversely polarized proton target was measured for the first time by the HERMES collaboration~\cite{ref:INT481}, ruling out the model of Ref.~\cite{ref:JJT98} and showing compatibility with predictions based on the spectator approximation~\cite{ref:noi06,ref:noi09} (see also the later Ref.~\cite{ref:INT486}). Preliminary results are available also from the COMPASS collaboration using deuteron~\cite{ref:Martin07} and proton~\cite{ref:INT487} targets. The 
$A^{\cos (\phi_R+\phi_{\bar{R}})}$ asymmetry in $e^+ e^-$ annihilation was recently measured by the BELLE collaboration~\cite{ref:BELLE}. By combining these data, the transversity $h_1$ was extracted for the first time in a collinear factorization scheme, properly including LO evolution effects in the behaviour of DiFF at different scales~\cite{ref:noiPRL}. 

\begin{figure}
\begin{center}
\includegraphics[width=0.7\textwidth]{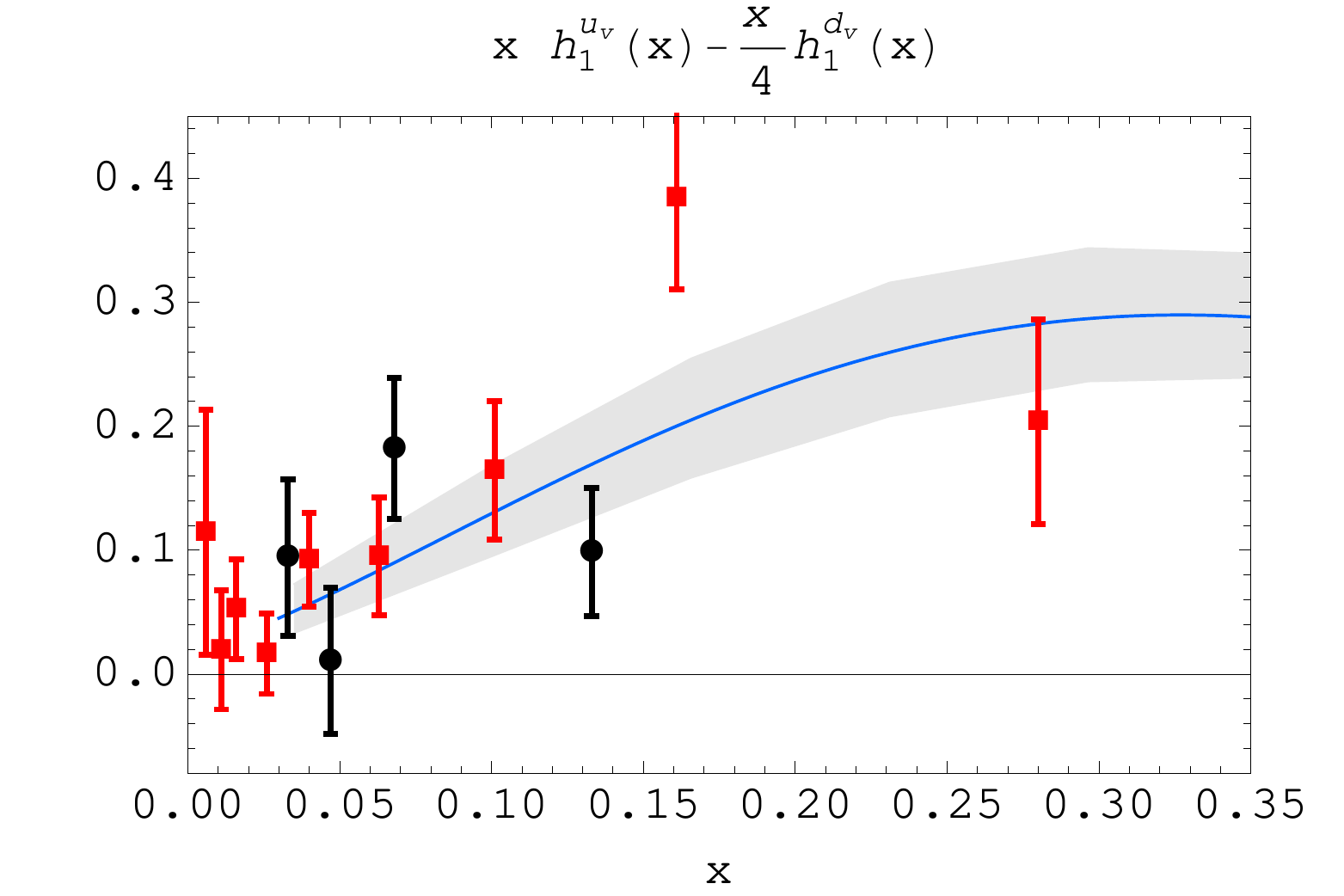}
\end{center}
\vspace{-0.5cm}
\caption{\label{fig:DiFFh1} The combination of valence $u,d$, flavors for transversity from 
Eq.~\ref{eq:DiFFh1}. Black circles for the SIDIS data from HERMES~\cite{ref:INT481}, red squares from COMPASS~\cite{ref:INT487}. The error bars are obtained by propagating the statistical errors of each term in the equation. The uncertainty band represents the same observable as deduced from the parametrization of Ref.~\cite{ref:TO09}.}
\end{figure}

The result is shown in Fig.~\ref{fig:DiFFh1}. By assuming certain symmetries of DiFF for $\pi^+ \pi^-$ pairs under isospin and charge conjugation transformations~\cite{ref:noi06,ref:noiPRL}, the SIDIS spin asymmetry $A_{UT}^{\sin (\phi_R+\phi_S) \sin\theta}$ can be written, after integrating upon 
$z, M_h^2$, as
\begin{eqnarray}
& &x \left[ h_1^{u_v} (x,Q^2) -\frac{1}{4} h_1^{d_v}(x,Q^2) \right]  =    \nonumber \\
& &\mbox{\hspace{1cm}} - 
\frac{A_{UT}^{\sin (\phi_R+\phi_S) \sin\theta}(x,Q^2)}{C_y}\, \frac{n_u(Q^2)}{n_u^\perp (Q^2)} 
\sum_{q=u,d,s} \frac{e_q^2}{e_u^2} \, x f_1^{q+\bar{q}}(x,Q^2) \; , 
\label{eq:DiFFh1}
\end{eqnarray}
where $x$ is the fractional momentum carried by quarks, $C_y$ is the depolarization factor, 
$h_1^{q_v}=h_1^q-h_1^{\bar{q}}$, $f_1^{q+\bar{q}}=f_1^q+f_1^{\bar{q}}$, and
\begin{eqnarray}
n_u (Q^2) &= &\int dz dM_h^2 D_{1,u}^{\pi^+ \pi^-}(z,M_h^2,Q^2) \nonumber \\
n_u^\perp (Q^2) &= &\int dz dM_h^2 \frac{|{\bf R}|}{M_h} \, H_{1,u}^{\open\,\pi^+ \pi^-}(z,M_h^2,Q^2) \; .
\label{eq:DiFFdensity}
\end{eqnarray}

The densitiy $n_u^\perp$ ($n_u$) of $\pi^+ \pi^-$ pairs produced by a (un)polarized quark $u$ can be calculated by fitting the $(z,M_h^2)$ dependence of DiFF in the $e^+ e^-$ asymmetry 
$A^{\cos (\phi_R+\phi_{\bar{R}})}$, as described in Ref.~\cite{ref:noiPRL}. The integrals in 
Eq.~\ref{eq:DiFFdensity} are performed over the range $0.5\leq M_h\leq 1$ GeV, $0.2\leq z\leq 0.7$, considered by HERMES~\cite{ref:INT481}. Evolution effects at LO~\cite{ref:noiDGLAP} produce a reduction of $n_u^\perp / n_u$ by 92\%$\pm$ 8\% when moving from the BELLE scale $Q^2=100$ GeV$^2$ down to the HERMES $\langle Q^2 \rangle = 2.5$ GeV$^2$. Using the MSTWLO08 set of unpolarized parton distributions $f_1$~\cite{ref:MSTW}, the data points of Fig.~\ref{fig:DiFFh1} are obtained by inserting experimental data in $A_{UT}^{\sin (\phi_R+\phi_S) \sin\theta}$ of 
Eq.~\ref{eq:DiFFh1}. Black circles correspond to the HERMES measurement of 
Ref.~\cite{ref:INT481}, red squares to the  COMPASS one of Ref.~\cite{ref:INT487}. The error bars are obtained by propagating the statistical errors of each term in Eq.~\ref{eq:DiFFh1}, the major role being played by the experimental statistical errors on 
$A_{UT}^{\sin (\phi_R+\phi_S) \sin\theta}$. The uncertainty band represents the same observable as deduced from the parametrization of Ref.~\cite{ref:TO09}. From Fig.~\ref{fig:DiFFh1} we deduce that there is a substantial agreement between the two extractions of $h_1$ obtained from two independent methods. However, the statistical meaning of the displayed error bars is totally different from the uncertainty band of Ref.~\cite{ref:TO09}: in order to perform a meaningful comparison a more detailed analysis is needed~\cite{ref:noiPRL}.


\section{Outlooks}

There are several interesting ongoing developments in each of the fields touched in previous Sections. 

As for single-hadron fragmentations, we are rapidly moving towards a full NNLO analysis of evolution effects when connecting data for $D_{1,i}^h(z,Q^2)$ at different $Q^2$ 
scales~\cite{ref:Moch11,ref:Albino11}. For $h=K$, nonsinglet fragmentation functions for $K^\pm$ can be directly extracted from data in a model independent way~\cite{ref:Albino10}, the present limitation being due to weak constrains coming from a not enough large data set. The ultimate goal would be to construct a nonperturbative error for $D_{1,i}^h$ by comparing in a meaningful way the most reliable parametrizations (like HKNS, DSS and AKK08, described in Tab.~\ref{tab:D1param}), similarly to what is done for parton distributions. To do so, a common interface similar to LHAPDF is needed, while at present only the web site {\tt http://www.pv.infn.it/$\sim$radici/FFdatabase/} is available. The dependence of $D_{1,i}^h$ upon the transverse momentum of $h$ is probably the most developing field, since very few experimental inputs are available. New results have been recently obtained which open the door to a correct treatment of evolution effects for TMD nonperturbative soft functions in the context of a suitable factorization theorem~\cite{ref:RogAy11}. 

As for two-hadron fragmentations, first data from the BELLE collaboration were released very recently and much more work has to done in order to unravel dihadron fragmentation functions and to confirm the first results about the extraction of transversity. In particular, a full flavor separation is in order for the analysis of $e^+ e^- \to (\pi^+ \pi^-) (\pi^+ \pi^-) X$ data, enlarging the explored range in invariant mass of the pion pair to study more contributing resonances. Moreover, new data from the COMPASS collaboration have been released~\cite{ref:Braunhere}, which demand for a more refined analysis. Finally, some data were released from the PHENIX collaboration about the 
$pp^\uparrow \to (\pi^+ \pi^-) X$ process~\cite{ref:PHENIX} (and more data will probably become available from the STAR collaboration in the near future), which should help in separating the antiquark components of dihadron fragmentation functions and, consequently, of 
transversity~\cite{ref:noipp}. 


\acknowledgments
I thank Alessandro Bacchetta and Barbara Pasquini for many fruitful discussions during the preparation of this document. This work is partially supported by the Italian MIUR through the PRIN 2008EKLACK, and by the European Community through the Research Infrastructure Integrating Activity HadronPhysics2 (Grant Agreement n. 227431) under the 7th Framework Programme. 


\end{document}